\documentstyle[prl,aps]{revtex}
\renewcommand{\baselinestretch}{1}
\begin{document}
\title{``Assisted cloning'' and ``orthogonal-complementing'' of an unknown state}
\author{Arun Kumar Pati}
\address{SEECS, Dean Street, University of Wales, Bangor LL 57 1UT, UK}
\date{\today}
\maketitle
\def\ra{\rangle}
\def\la{\langle}
\def\ver{\arrowvert}
\begin{abstract}
We propose a protocol where one can exploit dual quantum and classical channels
to achieve perfect ``cloning'' and ``orthogonal-complementing'' of an
unknown state with a minimal assistance
from a state preparer (without revealing what the input state is). 
The first stage of the protocol requires usual teleportation and
in the second stage,  the preparer disentangles the left over entangled
states by a single particle measurement process and
communicates a number of classical bits (1-cbit per copy) to different parties
so that perfect copies and complement copies are produced. We discuss our
protocol for producing two copies and three copies (and complement copies)
using two and four particle entangled state and suggest how to generalise
this for N copies and complement copies using multiparticle entangled state.

\end{abstract}

\vskip .5cm

PACS           NO:     03.67.Hk, 03.67.-a, 03.65.Bz\\

email:akpati@sees.bangor.ac.uk\\

\vskip 1cm

%\newpage
%\begin{multicols}{2}

\par
Manipulation and extraction of quantum information are important task in the
ongoing field of information theory. One of the interesting question
raised by Wootters and Zurek \cite{wz} and Dieks \cite{dk} is whether
it is possible to copy a quantum state perfectly. It was found that linearity
of quantum theory does not allow us to do so. In recent years this question
has again become important. Especially, in the field of quantum communication
and quantum computing one needs to know the ability to process the quantum
information content of a system. In the literature one finds another proof of
a ``no-cloning theorem'' which is based only on unitarity of the evolution. This says
that deterministic cloning of non-orthogonal states is impossible \cite{yu,ayu,bar}.
Even though we can not perfectly copy an unknown state, nevertheless, the
possibility of producing approximate copies of a quantum state has been
considered by Bu\v{z}ek and Hillery \cite{bh}. 
They \cite{hb} have shown that if two non-orthogonal states are cloned there will
be errors introduced in the copies. Bu\v{z}ek {\em et al} \cite{bbhb}
have discussed a network  for quantum copying where one can realise a Universal
Quantum Copying Machine (UQCM).
Bru{\ss} et al \cite{brub} have constructed an
optimal state dependent cloning machine for producing two clones.
There has been a proposal by Duan and Guo \cite{dugu}
for probabilistic cloning of non-orthogonal quantum state using unitary operator and measurement
process. The
possibility of producing $M$ copies from a given $N$ input states
has been investigated by Gisin and Masar \cite{gm}.
The possibility of
obtaining asymmetric clones were investigated by Cerf \cite{cerf}.
The problem of state discrimination and exact cloning has been discussed by
Chefles and Barnett \cite{chef}. They have also discussed a network which
interpolates between deterministic (approximate) and probabilistic (exact)
cloning \cite{chefb} from a given $M$ initial states.
Recently,
we have asked a question if there exists a cloning machine, which can produce
a linear superposition of multiple clones and found that such an ideal machine
can not exist \cite{akp1}. However, we have proved that if the machine fails
some times it is possible to build a novel cloning machine, which can
produce linear superposition of multiple clones \cite{akp2}. We have also
argued that the deterministic and probabilistic cloning machines are special
cases of our novel cloning machine. This would find potential
application in quantum information processing  as it would provide a parallel
quantum operation over many clones.

Recently, another limitation to quantum information
processing has been found. This says that it is not possible to produce a conjugate
copy of an unknown state using either linearity or unitarity of the quantum
process \cite{akp}.
The process of orthogonal-complementing of an unknown state is creating a state
which is orthogonal to the unknown state.
Independently,
it is found that it is not possible to perfectly create an orthogonal-complement
state of an
unknown state, because the operation involves anti-unitary operator
which can not be implemented in the laboratory on an isolated system \cite{bhw}.
In the case of a qubit a complement state can be related
to a conjugate state by a rotation operator. Hence, these problems are identical
(upto a rotation operator).
The optimal manipulation of an unknown state with a
universal NOT gate has been investigated in \cite{bhw}. A related
question on spin flipping has been investigated by Gissin and Popescu \cite{gp}
who found that it is not possible to flip an arbitrary spin.

    Although, there have been immense theoretical ideas about how well one
can copy a quantum state, none of them seems to yield a perfect copy of the
input state. Here, we should mention that in the Duan-Guo \cite{dugu} type
of probabilistic cloning
machine one can produce perfect copies with some probabilities. The probabilities
of success obey certain inequalities which depend on the inner product of the input
states. However, in the present scheme the perfect copies are produced with
a probability independent of the input state. Moreover, there is no proof yet
that using unitary and reduction operations one can produce an orthogonal
complement copy of an unknown state.
Also, there is no hope that one can test all the existing theoretical ideas
experimentally where one can create a copy with some error.

 The purpose of this paper is to
investigate the possibility of copying and complementing an unknown state
perfectly using
resources like entangled states, Bell-state measurement, single particle
von Neumann measurement and classical
information. We go beyond the traditional cloning ideas that exist in the
literature.
The question we raise is this: Can we produce a perfect copy of an unknown
quantum state with a minimal assistance from the state preparer (call him
Victor). It turns out that with the help of Victor our protocol
(which allows unitary and non-unitary (measurements) operations) can produce
a perfect copy and a complement copy. In a strict sense a quantum cloning
machine is not supposed
to communicate with Victor because one would say that since Victor knows what the state
is he could create as many copies as he wants and send them
to the interested parties.
Alternately, Victor could send his recipe to some-one to prepare the 
state which requires double infinity of bits of information \cite{rj}
to be sent across a classical channel. 
But if for some reason Victor does not want to do that what can we do
at best? To be more practical, we pose the question: If Victor has only one
input state and does not want to send infinite of bits of information, then
without violating
any principles of quantum theory and {\it using quantum mechanical operations
can we go arround ``no-cloning theorem''?}  If Victor agrees to assist the parties
with a minimum help without revealing what the input state is, then what
we can do is described in
this paper. It is shown that instead of sending infinite bits of information
Victor can use the entangled state left after the teleportaion process
 \cite{cb} and send one classical bit to Alice to create either a copy or an
orthogonal-complement copy of the unknown state. Like in the teleportation process one
 qubit can be passed by sending two cbits and the remaining flows across the
 entanglement channel, similarly here (in the second stage of protocol),
  the infinite amount of bits of information can be
 passed to a distant site by just sending one cbit and remaining bits flow
 across the entanglement channel. This is a non-trivial observation in this
 context which must be remembered and should not be confused with the 
 issue of what Victor  could do in principle. (What a state preparer could
 do in principle is not our concern here, what he is doing in the present
 protocol is important.) The importance of this approach can be in those
 situations, where Victor is cut-off (one way) from the rest of the world and
 he can not
 send any quantum states but has a classical channel to send only cbits. An
 example could be, suppose Victor had owned a private company (``qubit company'')
 which produces qubits and
 was sending to interested parties. One day, his company is crashed and
 he has no resources to start his company again nor has he capitals.
 Then, if the other interested parties can send him one half of the particles
 from
 an entangled source, then they can be benefited from geting a copy or a complement copy
 of an unknown state after receiving 1 cbit from Victor. This way many parties
 can benefit from Victor, although he has lost his ``qubit company''. But his
 knowledge and communication of 1 cbit per copy is quite helpful in such scenarios. 
 The machine which produces a perfect copy and a complement copy
of an unknown state
with the minimal assistance, we call ``assisted-cloning {\it cum} complementing
machine'' (ACCM).  Here, by
``minimal assistance'' we mean any classical communication from Victor
to Alice or Bob  {\it without
telling what the unknown state is. }

    Suppose we have a pure input state $\ver\Psi \ra_1 \in {\cal H} = C^2$,
     which is a qubit.
An arbitrary qubit can be represented as

\begin{equation}
\ver \Psi \ra_1 = \alpha \ver 0 \ra_1 + \beta \ver 1 \ra_1,
\end{equation}
where we
can choose  $\alpha$ to be real and $\beta$ to be a complex number, in general.
This qubit can be represented by a point on a sphere $S^2$ (which is the
projective Hilbert space for any two-state system) with the help of two real
parameters $\theta$ and $\phi$, where $\alpha = \cos {\theta \over 2}$ and
$\beta = \sin {\theta \over 2} \exp (i \phi)$. Now Victor 
prepares the input state and sends it through a ``ACCM''
to produce two output states and later he can verify how good these copies are.
Let Alice and Bob share one half of the particles from an EPR source as in
the quantum teleportation protocol \cite{cb}. The EPR state for
the particles 2 and 3 is given by,

\begin{equation}
\ver \Psi^- \ra_{23} = {1 \over \sqrt 2}(\ver 01 \ra_{23} - \ver 10 \ra_{23}).
\end{equation}

Suppose Alice is in possession of 2 and Bob is in possession of 3. The
input state $\ver \Psi \ra_1$ is unknown to both Alice and Bob. The combined state of the
unknown state 1 and the EPR state 23 (after reexpressing in terms of Bell-basis of
particle 1 and 2)  is

\begin{eqnarray}
&& \ver \Psi \ra_{123} = \ver \Psi \ra_1 \otimes \ver \Psi^- \ra_{23} =
  -{1 \over 2}[\ver \Psi^+ \ra_{12} (\sigma_z) \ver \Psi \ra_3  \nonumber \\
&& + \ver \Psi^- \ra_{12} \ver \Psi \ra_3 + \ver \Phi^+ \ra_{12} (i\sigma_y) \ver \Psi \ra_3 
+ \ver \Phi^- \ra_{12} (-\sigma_x)\ver \Psi \ra_3 ],
\end{eqnarray}
where $\sigma $'s are usual Pauli spin matrices generating $SU(2)$ Lie algebra.
Once Alice performs a Bell-state measurement onto two-particle
state 1 and 2, the resulting state is a tensor product of an entangled state
between particle 1 and 2 and the outgoing state (teleported state up to a
rotation operator). Suppose the measurement outcome of Alice is
$\ver \Psi^- \ra_{12}$, (the probability of this outcome is only ${1 \over 4}$)
then the resulting three particle state is given by

\begin{equation}
\ver \Psi^- \ra_{12} \la \Psi^- \ver \Psi \ra_{123} =  -{1 \over 2} \ver \Psi^- \ra_{12} \otimes \ver \Psi \ra_{3}.
\end{equation}

     When Alice communicates her measurement results to Bob (classical two bits
of information) then Bob knows that he has received the original state. If the
measurement outcome of Alice is other than $\ver \Psi^- \ra_{12}$, then Bob
has to apply suitable rotation operator to get the original state.
Now can we create a copy of the original state at Alice's place?
Since the teleportation process obeys  the ``no-cloning  theorem'',
at first instance
it might look impossible to have a copy at Alice's place. However, this impossibility
becomes a possibility when we allow {\it one bit of classical information} from
Victor to Alice.

Usually it is said that the Bell-state measurement destroys the particle 1
and 2, so there is no state available to Alice. This is some times
stressed emphatically that the quantum teleportation process obeys ``no-cloning''
theorem, so particle 1 is destroyed. But careful analysis shows that obeying
the  ``no-cloning theorem''  of quantum theory and making two-particle detection
 (Bell-state measurements) are two different things. The point to be noted
 is that, in general
the particles 1 and 2 need not be destroyed after Bell-state measurement. It
so happens that when one uses photons for Bell-state analysis as
was done by Bouwmeester et al \cite{dik} they are absorbed
by the detectors and hence destroyed. For other particles this need not be so.
Because the projection postulate says that when we apply a Bell-basis
projector onto a
the combined state $\ver \Psi \ra_{123}$ indeed a Bell-state remains formally
(see eqn.(4)).
Quantum theory does not say that the particle $1$ and $2$ need
to be destroyed after a measurement.
How to implement this in practice is another question. A possible way to do
this is
through Bell-state analysis using quantum-nondemolition measurement (QND)
\cite{bvt} in the case of photons. This would need a device which can perform
photon number QND measurement at single photon level. By sending particles
1 and 2 along with probe modes in a nonlinear Kerr medium one may think
of a QND measurement. This would be, indeed, a challenge for experimentalist
in the future. For other entangled sources such as spin-1/2
particles or two-level atoms one needs some suitable detector which can
distinguish all four Bell-states and allow them to propagate freely
for further processing.
Let there be particles 1 and 2 after Bell-state measurement, which are
in a singlet state. Alice
sends particle 1 to Victor and keeps particle 2 in her possession.
Now can Victor do some thing (without really telling what the input state is)
so that Alice gets a copy of the unknown state? We show that this is possible
in principle. Victor carries out another measurement on particle
1 by putting a linear polariser (in case of photon) or Stern-Gerlach apparatus
(in case of spin 1/2 particles) to measure the state in another basis $\{\ver x \ra,
\ver y \ra\}$, where the new basis is related to the old basis $\{\ver 0 \ra, \ver 1 \ra \}$
in the following manner

\begin{eqnarray}
\ver 0 \ra_1 = \alpha \ver x \ra_1 + \beta \ver y \ra_1 \nonumber\\
\ver 1 \ra_1 = \beta^* \ver x \ra_1 - \alpha \ver y \ra_1.
\end{eqnarray}

Notice that the normalisation and orthogonality relation between basis
vectors are preserved
under this transformation. Interestingly, we find that the basis $ \ver x \ra_1
 = \ver \Psi \ra_1$
and the basis $\ver y \ra_1 = \ver \Psi_{\perp} \ra_1$, where
$\ver \Psi_{\perp} \ra_1 = (\alpha \ver 1 \ra_1 - \beta^* \ver 0 \ra_1)$
is the orthogonal-complement state to $\ver \Psi \ra_1$. However,
we keep  $ \ver x \ra_1 , \ver y \ra_1$ for Victor just to distinguish
the fact that he knows the state. When we write $ \ver \Psi \ra$
and $\ver \Psi_{\perp} \ra$ for other particles we mean they are unknown
to the parties concerned.
Now writing the entangled state $\ver\Psi^- \ra_{12}$ in
the basis $ \ver x \ra_1 , \ver y \ra_1$ gives us 

\begin{eqnarray}
\ver \Psi^- \ra_{12} = {1 \over \sqrt 2}[ \ver x \ra_1 (\alpha \ver 1 \ra_2 -
\beta^* \ver 0 \ra_2) + \ver y \ra_1 (\alpha \ver 0 \ra_2 + \beta \ver 1 \ra_2) ].
\end{eqnarray}
If the outcome of Victor is $\ver y \ra_1$, then he sends his measurement result
(one bit of classical information)
to Alice. Alice knows that her state of particle 2 has been found in the
original state   $(\alpha \ver 0 \ra_2 + \beta \ver 1 \ra_2)$
which is just a copy.

    More explicitly, the total state after a Bell-state measurement
and a single particle von-Neumann measurement is given by
 
\begin{equation}
\ver y \ra_1 \la y\ver \Psi^- \ra_{12} \la \Psi^- \ver \Psi \ra_{123} =  -{1 \over 2 \sqrt 2}
 \ver y \ra_1 \otimes \ver \Psi \ra_{2} \otimes \ver \Psi \ra_{3}.
\end{equation}

If the outcome of Victor's measurement result is $\ver x \ra_1$ then the classical
communication from Victor would tell Alice that she has obtained a state
which is $ (\alpha \ver 1 \ra_2 - \beta^* \ver 0 \ra_2) $. This is a
complement copy of the unknown state. Now the  state (if the outcome is
$\ver x \ra_1$) is given by
 
\begin{equation}
\ver x \ra_1 \la x\ver \Psi^- \ra_{12} \la \Psi^- \ver \Psi \ra_{123} =  -{1 \over 2 \sqrt 2}
 \ver x \ra_1 \otimes \ver \Psi_{\perp} \ra_{2} \otimes \ver \Psi \ra_{3}.
\end{equation}

   On the other hand, if the measurement outcome of Alice is other
than $\ver \Psi^- \ra_{12}$ in the first stage of the protocol, then the result
of the second stage of the protocol can be worked out in detail. First, we note that though
the singlet state is same in any basis, the other Bell-states are not the same.
In the basis $\{\ver x \ra_1, \ver y \ra_1 \}$ we can express the
other three Bell-states as

\begin{eqnarray}
&& \ver \Psi^+ \ra_{12} = - {1 \over \sqrt 2}[ \ver x \ra_1 (\sigma_z) \ver \Psi_{\perp} \ra_{2} +
\ver y \ra_1 (\sigma_z) \ver \Psi \ra_{2} ]    \nonumber \\
&& \ver \Phi^+ \ra_{12} =  {1 \over \sqrt 2}[ \ver x \ra_1 (i\sigma_y) \ver \Psi_{\perp} \ra_{2} +
\ver y \ra_1 (i\sigma_y) \ver \Psi \ra_{2} ]    \nonumber \\
&& \ver \Phi^- \ra_{12} =  {1 \over \sqrt 2}[ \ver x \ra_1 (\sigma_x) \ver \Psi_{\perp} \ra_{2} +
\ver y \ra_1 (\sigma_x) \ver \Psi \ra_{2} ]   
\end{eqnarray}

When Alice's outcome is $\ver \Psi^+ \ra_{12}$ in the first stage of the protocol,
then the resulting states
after a Bell-state measurement, a single particle von Neumann measurement
and classical communication are given by

\begin{eqnarray}
&& \ver x \ra_1 \la x\ver \Psi^+ \ra_{12} \la \Psi^+ \ver \Psi \ra_{123} =  {1 \over 2 \sqrt 2}
 \ver x \ra_1 \otimes (\sigma_z)\ver \Psi_{\perp} \ra_{2} \otimes (\sigma_z)\ver \Psi \ra_{3} \nonumber \\
&& \ver y \ra_1 \la y\ver \Psi^+ \ra_{12} \la \Psi^+ \ver \Psi \ra_{123} = {1 \over 2 \sqrt 2}
 \ver y \ra_1 \otimes (\sigma_z)\ver \Psi \ra_{2} \otimes (\sigma_z)\ver \Psi \ra_{3}.
\end{eqnarray}
Similarly, 
when Alice's outcome is $\ver \Phi^+ \ra_{12}$, then the resulting states
after a Bell-state measurement, a single particle von Neumann measurement
and classical communication are given by

\begin{eqnarray}
&& \ver x \ra_1 \la x\ver \Phi^+ \ra_{12} \la \Phi^+ \ver \Psi \ra_{123} = - {1 \over 2 \sqrt 2}
 \ver x \ra_1 \otimes (i\sigma_y)\ver \Psi_{\perp} \ra_{2} \otimes (i\sigma_y)\ver \Psi \ra_{3} \nonumber \\
&& \ver y \ra_1 \la y\ver \Phi^+ \ra_{12} \la \Phi^+ \ver \Psi \ra_{123} = -{1 \over 2 \sqrt 2}
 \ver y \ra_1 \otimes (i\sigma_y)\ver \Psi \ra_{2} \otimes (i\sigma_y)\ver \Psi \ra_{3}.
\end{eqnarray}
Finally, 
when Alice's outcome is $\ver \Phi^- \ra_{12}$, then the resulting states
after a Bell-state measurement, a single particle von Neumann measurement and
classical communication are given by

\begin{eqnarray}
&& \ver x \ra_1 \la x\ver \Phi^- \ra_{12} \la \Phi^- \ver \Psi \ra_{123} =  {1 \over 2 \sqrt 2}
 \ver x \ra_1 \otimes (\sigma_x)\ver \Psi_{\perp} \ra_{2} \otimes (\sigma_x)\ver \Psi \ra_{3} \nonumber \\
&& \ver y \ra_1 \la y\ver \Phi^- \ra_{12} \la \Phi^- \ver \Psi \ra_{123} = {1 \over 2 \sqrt 2}
 \ver y \ra_1 \otimes (\sigma_x)\ver \Psi \ra_{2} \otimes (\sigma_x)\ver \Psi \ra_{3}.
\end{eqnarray}
Thus, (7),(8) and (10-12) are exact ones and the main results of this paper.
Our analysis 
illustrates the power of Bell-state and single particle measurements. The important
observation is that if after the Bell-state measurement Victor finds $\ver y \ra_1$
in a single particle measurement, then 1-cbit from Victor to Alice will result
in an exact copy or a copy up to a rotation operator at Alice's place. If Victor
finds $\ver x \ra_1$, then sending of 1-cbit will yield an exact replica of
complement state or a complement state up to a rotation operator. Interestingly,
the rotation operators that Alice has to apply to get a copy is same as that
of the Bob's case to get the original state.

In the special case, if we restrict our unknown state to be real, i.e.,
$\ver \Psi \ra = \cos \theta \ver 0 \ra + \sin \theta \ver 1 \ra $, which means
on the projective Hilbert space $S^2$ the point lies on the equatorial line,
then the azimuthal angle is zero.  Alice just has to perform a rotation or do
nothing after
receiving the classical information from Victor in case the measurement
outcome is  $\ver x \ra_1$ or $\ver y \ra_1$ and in both cases she gets
a copy of the unknown state. Thus for a real unknown state our protocol
produces a clone $100 \%$ of the time.
For an arbitrary unknown state this protocol is able to produce an
accurate copy of the original
input state $50 \%$ of the time and an orthogonal-complement copy $50\%$ of the time.
Had our protocol worked only in the case of $\ver \Psi^- \ra$, then
the probability to produce a clone would have been ${1 \over 8}$. But in the present
case our protocol works for all Bell-state outcomes, hence probability to
produce a copy is ${1 \over 2}$, as mentioned.
The fidelity of getting the copy and the complement copy is {\em maximum, i.e. unit.}
One may tend to think that since the preparer is
communicating with Alice this protocol may not qualify to be called quantum
cloning. However, note that  although Victor communicates
the measurement
result on some basis he is not going to tell what the exact value of
$\alpha$ and $\beta$
is. The measurement result sent by Victor does not contain all
information about the unknown state. In fact, the state of the particle 1 is no
longer the original one, it is a member of the entangled pair. Knowledge of
the unknown state is still not revealed to Alice and Bob. So this
is not a classical cloning machine.

      Can we imagine producing more copies or complement copies using
the ``assisted-quantum cloning'' and  ``complementing''
protocol? That is, given an input state $\ver \Psi \ra$ can we have an output state
$\ver \Psi \ra \otimes \ver \Psi \ra \otimes \ver \Psi \ra$  or more?  At first
sight it may
seem that may be if we use three particle entanglement we could generate
{\it one to three}
copies at different  sites with the help of Victor. But it turns out that
with a three-particle entangled source one can again produce only one perfect
copy or one complement copy. So this analysis we are not going to present. The useful resource for
producing {\it one to three} copies or 2 copies and a complement copy
is a four-particle entangled state of the type
Greenberger-Horne-Zeilinger (GHZ) \cite{ghz}. This four particle state was
originally proposed in the context of decay of four spin-half particles and
was shown to yield non-local correlations without inequalities. The four
particle entangled state (GHZ state) is given by

\begin{equation}
\ver \Psi \ra_{2345} = {1 \over \sqrt 2}(\ver 0011 \ra_{2345} + \ver 1100 \ra_{2345} )
\end{equation}

Suppose we have three parties Alice, Bob and Carla who share the particles
from this entangled
state. Alice has particle 2, Bob has particles 3 and 4 and Carla has
particle 5. Let Victor prepare a state which is unknown to Alice, Bob and Carla
and is given by (1). He sends particle 1 to Alice who is now in possession of
two particles 1 and 2. Now the combined five particle state is given by
$\ver \Psi \ra_1 \otimes \ver \Psi \ra_{2345}$. Let us express the basis of
states of particle
1 and 2 in their respective Bell-basis. Then the total state can be written as

\begin{eqnarray}
&& \ver \Psi \ra_{12345} = \ver \Psi \ra_1 \otimes \ver \Psi \ra_{2345} 
{1 \over 2}[\ver \Psi^+ \ra_{12} (\beta \ver 011 \ra_{345} - \alpha \ver 100 \ra_{345})  \nonumber\\
&& - \ver \Psi^- \ra_{12} ( \beta \ver 011 \ra_{345}  + \alpha \ver 100 \ra_{345}) 
+ \ver \Phi^+ \ra_{12} (\alpha \ver 011 \ra_{345} - \beta \ver 100 \ra_{345}  )  \nonumber\\
&& + \ver \Phi^- \ra_{12} ( \alpha \ver 011 \ra_{345}  + \beta \ver 100 \ra_{345})
\end{eqnarray}

Now Alice carries out a Bell-state measurement on the particle 1 and 2. The
two-particle projection would yield any one of four possible results. If the
readout of Alice's measurement is $\ver \Psi^- \ra_{12}$ then the resulting state
will be

\begin{equation}
\ver \Psi^- \ra_{12} \la \Psi^- \ver \Psi \ra_{12345} = 
-{1 \over 2}
 \ver \Psi^- \ra_{12} ( \beta \ver 011 \ra_{345}  + \alpha \ver 100 \ra_{345})
\end{equation}
Thus the Bell-state measurement leaves particle 1 and 2 in a maximally entangled
state and  particle 3,4 and 5 in a three-particle entangled state. After the above
measurement
Alice sends her results via classical channel two bits of information to both
Bob and Carla. In next step, Bob who is in possession of particle 3 and 4,
carries out another Bell-state measurement  on them. Let us express the
above state in terms of Bell-states of 3 and 4. This is given by

\begin{eqnarray}
&& \ver \Psi^- \ra_{12} \la \Psi^- \ver \Psi \ra_{12345} = 
-{1 \over 2}
 \ver \Psi^- \ra_{12} [\ver \Psi^+ \ra_{34}(\alpha \ver 0 \ra_{5} + \beta \ver 1 \ra_{5}) \nonumber \\
&& - \ver \Psi^- \ra_{34}(\alpha \ver 0 \ra_{5} - \beta \ver 1 \ra_{5}) ] 
\end{eqnarray}
After a Bell-state measurement onto particle 3 and 4 if the readout is
(say) $\ver \Psi^+ \ra_{34}$,
then the state of particle 5 is found to be in the original state. If the read out is
$\ver \Psi^- \ra_{34}$ then the state of particle 5 is nothing but the original
state up to a rotation operator $\sigma_z$. Suppose Bob's measurement gives a
result $\ver \Psi^+ \ra_{34}$, then the classical information from Bob to Carla
would yield the state

\begin{equation}
\ver \Psi^+ \ra_{34} \la \Psi^+ \ver \Psi^- \ra_{12} \la \Psi^- \ver \Psi \ra_{12345} = 
-{1 \over 2}
 \ver \Psi^- \ra_{12} \otimes \ver \Psi^+ \ra_{34} \otimes \ver \Psi \ra_{5}  
\end{equation}

Note that in the second stage Bob needs to send only one bit of classical
information to Carla as he could get only two possible Bell-state measurement
results. The resulting state at Carla's hand is nothing but the teleported state
of the original input state (this can be called teleportation of an unknown state
using four-particle entangled state, which has not been discussed in the literature).
Recently, we find Karlsson and Bourennane \cite{kb} have discussed teleportation
of an unknown state using three particle entangled state. But as we have said,
three-particle entanglement is not a useful resource for cloning and
complementing unknown states.
After teleportation of the original state the particle 1 and 2 and 3 and 4 are
in a maximally entangled state. Now Alice and Bob send particle 1 and 3 to Victor
one after the other. When Victor gets the particles (1 and 3) he chooses to
measure the states in the basis $\ver x \ra_i, \ver y \ra_i, (i = 1,3)$, where

\begin{eqnarray}
\ver 0 \ra_i = \alpha \ver x \ra_i + \beta \ver y \ra_i \nonumber\\
\ver 1 \ra_i = \beta^* \ver x \ra_i - \alpha \ver y \ra_i
\end{eqnarray}

In the new basis the total state is given by

\begin{eqnarray}
&&\ver \Psi^+ \ra_{34} \la \Psi^+ \ver \Psi^- \ra_{12} \la \Psi^- \ver \Psi \ra_{12345} =  \nonumber\\
&& {1 \over 4}[ \ver x \ra_3 (\alpha \ver 1 \ra_4 +
\beta^* \ver 0 \ra_4) - \ver y \ra_3 (\alpha \ver 0 \ra_4 - \beta \ver 1 \ra_4) ]
\otimes  \nonumber\\
&&[ \ver x \ra_1 (\alpha \ver 1 \ra_2 -
\beta^* \ver 0 \ra_2) + \ver y \ra_1 (\alpha \ver 0 \ra_2 + \beta \ver 1 \ra_2) ]
\otimes \ver \Psi \ra_5
\end{eqnarray}

Suppose Victor first performs a von Neumann measurement on particle 1 and then
on 3 and in both cases let the
outcomes be $\ver y \ra_1$ and $\ver y \ra_3$. He can send the classical
information ( one cbit ) to  Alice and (one  cbit) to Bob who can in turn
find their particles in the original state exactly or up to a rotation operator,
respectively. Thus the final state after von Neumann measurements is given by

\begin{eqnarray}
&& \ver y \ra_1 \la y \ver y \ra_3 \la y \ver \Psi^+ \ra_{34} \la \Psi^+ \ver
 \Psi^- \ra_{12} \la \Psi^- \ver \Psi \ra_{12345}  =  \nonumber \\
&&  -{1 \over 4}
 \ver y \ra_1 \otimes \ver \Psi \ra_{2} \otimes \ver y \ra_3
 \otimes (\sigma_z) \ver \Psi \ra_{4} \otimes \ver \Psi \ra_{5}.
\end{eqnarray}

Since particles 2, 4, and 5 are in possession of Alice, Bob and Carla,
respectively it is clear from (20) that each of them acquire a perfect copy
of the unknown state.
On the other hand if Victor's outcome for particle 1 and 3 are $\ver x \ra_1$
and $\ver x \ra_3$, then the resulting state after communicating one bit
to Alice and one to Bob is given by

\begin{eqnarray}
&& \ver x \ra_1 \la x \ver x \ra_3 \la x \ver \Psi^+ \ra_{34} \la \Psi^+ \ver \Psi^- \ra_{12} \la \Psi^- \ver \Psi \ra_{12345}
 =  \nonumber \\
&&  -{1 \over 4}
 \ver x \ra_1 \otimes \ver \Psi_{\perp} \ra_{2} \otimes \ver x \ra_3 \otimes
  (-\sigma_z) \ver \Psi_{\perp} \ra_{4} \otimes \ver \Psi \ra_{5}.
\end{eqnarray}
In this case Alice gets a complement copy, Bob gets a complement copy (up to
a rotation operator) and Carla gets the original state.
Similarly, depending on the outcomes of the measurement process on  particle
1 and 3 two  other possibilities exist. In general when Victor finds both
particles in the basis $\ver y \ra$, then we have $2$ copies with probability
$1/16$ and when both particles are found in the basis $\ver x \ra$, we have
$2$ complement copies with probability $1/16$. However, if Victor finds particle
$1$ in the basis $\ver x \ra$ and $3$ in $\ver y \ra$ (or vice versa) then we
have a copy and a complement copy (all are up to doing nothing or a rotation
operation) with probability $1/8$. 
Thus the above protocol is able to produce two perfect copies or two
complement copies or a copy and a complement copy (including original one
) of an unknown state with the help of Victor. Since our protocol would
work for all Bell-state outcomes, therefore, the probability of producing 2
clones (upto a rotation operator) is $25 \%$ and of producing 2 complement
copies is $25 \%$ and 1 clone and 1 complement is $50 \%$.
It is a curious interplay between dual quantum and classical channel which
is surprisingly enough able to produce {\it one to two} and
{\it one to three} perfect copies and complement copies. 
The classical information
(1 cbit) send by Victor is not enough to create a perfect copy from some arbitrary state
{\it unless they have shared one half of the particle from the entangled state.}
Here, again left-over entanglement after the teleportation process plays an
important role in sending infinite bits from a state preparer. The present
results can be generalised to produce multiple copies and complement copies
using multiparticle entanglement source and multiparties. One can ask what is
the useful resource 
to produce $N$ copies or $N$ complement copies  (plus 1 original). It
can be argued that we need
$n = 2N$ particle entanglement source and $N+1$ parties to be involved.
The entangled source has to be distributed in such a way that the first
 and the last
person possess one particle each and all intermediate parties possess 2
particles from the entangled source.

   To conclude this paper, we have proposed a protocol which can produce
perfect copies and orthogonal-complement copies of an unknown state with the help of
dual quantum and classical
channel (including minimal communication from state preparer). Though, in principle
the state preparer knows what the state is he never reveals the information
about the unknown state. This protocol uses the minimal amount of additional
information (1 cbit) to produce a copy or a complement copy.
This is the first time we are able to produce
perfect copies and complement copies of an unknown state within 
quantum theory
(which takes into account measurement processes).
It is an open problem in quantum
information theory how can one produce a copy or a complement copy of an
unknown state within a quantum computer, where
it is highly desirable to copy the  information again and again. At least, we
have shown that it is possible to do so perfectly with a minimal assistance
from the state preparer. Our protocol provides an insight on what extra
resource do we need to achieve perfect cloning and complementing. We hope
this will be a practical way of copying and
complementing quantum states in future. Also, the present work could have
some application in quantum communication complexity \cite{kn} and in quantum bit
commitment protocols \cite{hklo}.  

\vskip 1cm
 I wish to thank S. Braunstein useful discussions. I thank V. Bu\v{z}ek and
 A. Chefles for going through my
 paper and sending their comments at some stage or the other. I thank L. M. Duan
 for useful discussions concerning the ``no-cloning'' theorem and teleportation
 process. I wish to thank D. Dieks for encouragements. Also, I thank P. Kok for
 a careful reading of the manuscript. I gratefully acknowledge the financial support from European Physical Science
 Research council (EPSRC).

\renewcommand{\baselinestretch}{1}

%\end{multicols}

\end{document}